\documentclass[sn-standardnature]{sn-jnl}

\usepackage{fancyvrb}

\jyear{2023}%

\theoremstyle{thmstyleone}%
\theoremstyle{thmstyletwo}%

\theoremstyle{thmstylethree}%
\begin{document}

\title[The 21-cm signal from the dark ages]{Prospects for precision cosmology with the 21 cm signal from the dark ages}


\author[]{\fnm{Rajesh} \sur{Mondal}}\email{mondalr@tauex.tau.ac.il}

\author[]{\fnm{Rennan} \sur{Barkana}}\email{barkana@tauex.tau.ac.il}
\equalcont{These authors contributed equally to this work.}

\affil[]{\orgdiv{School of Physics and Astronomy}, \orgname{Tel Aviv University}, 
\city{Tel Aviv}, 
\postcode{69978}, \country{Israel}}


\abstract{The 21 cm signal from the dark ages provides a potential new probe of fundamental cosmology. While exotic physics could be discovered, here we quantify the expected benefits within the standard cosmology. A measurement of the global (sky-averaged) 21 cm signal to the precision of thermal noise from a 1,000 h integration would yield a measurement within 10\% of a combination of cosmological parameters. A 10,000 h integration would improve this measurement to 3.2\% and constrain the cosmic helium fraction to 9.9\%. Precision cosmology with 21 cm fluctuations requires a collecting area of 10 km$^2$ (corresponding to 400,000 stations), which, with a 1,000 h integration, would exceed the same global case by a factor of $\sim2$. Enhancing the collecting area or integration time by an order of magnitude would yield a 0.5\% parameter combination, a helium measurement five times better than Planck and a constraint on the neutrino mass as good as Planck. Our analysis sets a baseline for upcoming lunar and space-based dark-ages experiments.}






\maketitle

Observation of the redshifted 21-cm signal due to the hyperfine transition of neutral hydrogen\,(HI) is a promising method to study its three-dimensional\,(3D) distribution in the Universe \cite{Sunyaev1972, Hogan, Scott1990}. The era from the epoch of recombination (redshift $z\sim1100$), when matter decoupled from the radiation and these photons free-streamed as the cosmic microwave background\,(CMB), until the formation of the first substantial population of luminous objects ($z\sim 30$) is referred to as the `Dark Ages'. After decoupling, the temperature of the gas\,($T_{\rm g}$) declined adiabatically as $(1+z)^2$ whereas the CMB temperature\,($T_{\gamma}$) fell as $(1+z)$. The spin temperature $T_{\rm s}$ (an effective temperature that measures the ratio of the population densities of the upper and lower states of the 21-cm transition) was strongly coupled to $T_{\rm g}$ through collisional coupling until $z\sim70$~\cite{madau97}. After this time, the collisional process became ineffective and $T_{\rm s}$ began to approach $T_{\gamma}$. Thus, during the dark ages $T_{\rm s}$ remained significantly lower than $T_{\gamma}$ over a wide redshift range of $300 \lesssim z \lesssim 30$, during which the HI is expected to produce absorption features in the CMB spectrum.

The dark ages are potentially a critically important window in the evolutionary history of the Universe. Since cosmic evolution at this time was not yet significantly affected by astrophysical processes, the dark ages offer a clean probe of fundamental cosmology similar to the CMB. This is in contrast to cosmological probes in the modern Universe, such as those based on the galaxy distribution or on the statistics of Lyman-$\alpha$ absorption lines, that suffer an irreducible systematic uncertainty due to the potential influence of complex astrophysics (including star formation, radiative feedback from stars and stellar remnants, and supernova feedback). The dark ages can be probed using the redshifted HI 21-cm signal, either by measuring the global (or mean) signal or by measuring the fluctuations at various length scales, i.e., the power spectrum. Unlike the CMB which comes to us from a single cosmic time, the 21-cm signal can be observed over a range of cosmic times, as each frequency 1420\,MHz$/(1+z)$ corresponds to a different look-back time. The 21-cm data set is thus 3D; moreover, since small-scale fluctuations that are washed out in the CMB are available in the 21-cm power spectrum, the latter contains potentially far more cosmological information~\cite{Loeb2004}.

Prior to recombination, the coupling of the baryons to the photons kept the baryon density and temperature fluctuations negligible on sub-horizon scales. The 21-cm power spectrum during the dark ages probes the era of baryonic infall into the dark matter potential wells~\cite{barkana2005} and is quite sensitive to the values of the $\Lambda$CDM cosmological parameters~\cite{barkana2005}, particularly on the scale of the baryon acoustic oscillations (BAO) that are a remnant of the pre-recombination baryon-photon~fluid. Importantly, the fluctuations on these scales are still quite linear during the dark ages, and thus modeling them does not need to deal with complex non-linearity as is the case for probes in the more recent Universe. The 21-cm fluctuations probe fluctuations of the baryon density, peculiar velocity \cite{bharadwaj04,barkana05}, and baryon temperature (as determined by the fluctuating sound speed \cite{naoz05,barkana2005}). A number of smaller contributions must be included in a precise calculation~\cite{Lewis2007, Ali-Ha2014}. All of this assumes the standard cosmology. Besides this, there are various studies on non-standard possibilities during the dark ages, such as DM-baryon interactions~\cite{Tashiro14, Munoz2015, Barkana2018} or features in the primordial power spectrum~\cite{2016JCAP}. It has also been shown that the 21-cm signal can potentially be a powerful probe of primordial non-Gaussianity, at the levels expected from cosmic inflation (see, e.g., \cite{Loeb2004,Pillepich2007, Joudaki2011, Floss2022, Balaji2022}), but below we find that it would be difficult observationally to reach the high wavenumbers where this is most promising. While some types of exotic (non-standard) cosmology would be easier to detect, we focus in this work on the safest case of standard cosmology.

Observing the 21-cm signal from the dark ages using radio telescopes on Earth would be nearly impossible due to the ionosphere, which heavily distorts and eventually blocks very low frequencies. This necessitates lunar or space-based experiments, and these are being rapidly developed as part of the international race to return to the moon, with efforts including NCLE ({\url{https://www.ru.nl/astrophysics/radboud-radio-lab/projects/netherlands-china-low-frequency-explorer-ncle}}) (Netherlands-China), DAPPER ({\url{https://www.colorado.edu/project/dark-ages-polarimeter-pathfinder}}) (USA), FARSIDE ({\url{https://www.colorado.edu/project/lunar-farside}}) (USA), DSL ({\url{https://www.astron.nl/dsl2015}}) (China), PRATUSH ({\url{https://wwws.rri.res.in/DISTORTION/pratush.html}}) (India), FarView ({\url{https://www.colorado.edu/ness/projects/farview-lunar-far-side-radio-observatory}}) (USA), SEAMS (India), LuSee Night ({\url{https://www.lusee-night.org/night}}) (USA), ALO ({\url{https://www.astron.nl/dailyimage/main.php?date=20220131}}) (Europe), and ROLSES ({\url{https://www.colorado.edu/ness/projects/radiowave-observations-lunar-surface-photoelectron-sheath-rolses}}) (USA). These missions will probe either the global signal or the spatial fluctuations of the dark ages 21-cm signal. We note that most of these experiments are at the early design stage, and also measuring the dark ages power spectrum is substantially more futuristic than the global signal.

Given the great potential for precision cosmology, as well as the rapid observational developments, in this paper we study the use of the 21-cm signal (both the global signal and power spectrum) during the dark ages to constrain the cosmological parameters. In this prediction it is important to account for observational limitations, and not just assume the theoretical limiting case of a full-sky, cosmic variance limited experiment. Indeed, an inevitable obstacle is thermal noise, which rises rapidly with redshift as well as wavenumber. For the power spectrum, another observational barrier is the angular resolution, since  for an interferometer with a given set of antennae, reaching a higher angular resolution reduces the array's sensitivity. In general in 21-cm cosmology, an interferometer requires a much greater investment of resources than a simple antenna for measuring the global signal; the potential reward of the former is also substantially greater due to the richer information content available in measuring the power spectrum versus redshift. We note that the just-mentioned observational obstacles worsen more rapidly with redshift for the 21-cm power spectrum; the signal also declines faster for the power spectrum, since the dark ages universe was more homogeneous, density fluctuations were smaller, and there were no galaxies around to amplify the 21-cm fluctuations. These factors make the global signal relatively advantageous at least as an initial probe of the dark ages.

There are other practical difficulties faced by 21-cm experiments, including removing terrestrial radio-frequency interference (RFI), accounting for the effect of the ionosphere, and removing or avoiding foreground emission (coming from synchrotron radiation from our own Milky Way as well as other galaxies). As these obstacles are increasingly overcome, this will allow for deeper integrations for which the noise remains dominated by the thermal noise. Techniques for achieving this are a matter of research that is achieving continuous improvement, as reflected in the best current constraints from global experiments such as EDGES~\cite{Bowman:2018} and SARAS~\cite{SARAS}, and interferometers such as LOFAR~\cite{LOFAR-EoR:2020}, MWA~\cite{Trott:2020} and HERA~\cite{HERA}. Thus, for the 21-cm signal from the dark ages, the thermal noise for various integration times serves as a fiducial benchmark for future experiments. Note that going to the moon could present substantial practical advantages beyond just avoiding the Earth's ionosphere: a potentially benign environment that is extremely dry and stable, plus the blocking out of terrestrial RFI (on the lunar far-side).


\section*{Calculating the 21-cm signal}
\label{sec:signal}
As previously noted, the 21-cm differential brightness temperature relative to the CMB, ${T}_{\rm b}$, from each redshift $z$ is observed at a wavelength of $21 \times (1 + z)$ cm. Global experiments measure the cosmic mean 21-cm brightness temperature; since these are relatively simple and are advantageous at the highest redshifts, we consider them first in Sec.~\ref{sec:global}. In addition, the 21-cm brightness temperature fluctuates spatially, mainly due to the fluctuations in the gas density and temperature. As noted above, the gas retains some memory of the early BAO, on the scale traversed by sound waves in the photon-baryon fluid (wavenumber $k \sim 0.1\,{\rm Mpc}^{-1}$; scales are comoving unless indicated otherwise). The signature of these oscillations can be detected in the 21-cm power spectrum from the dark ages (see Sec.~\ref{sec:power}). 

We use the standard \texttt{CAMB} ({\url{http://camb.info}}) \cite{Lewis2007,CAMB} cosmological perturbation code to precisely generate the 21-cm global signal from the dark ages and, after accounting for the anisotropy due to redshift space distortions~\cite{BLlos}, the 3D 21-cm power spectrum. While the line-of-sight anisotropy is in principle measurable, foreground removal is expected to make this difficult, so here we conservatively consider only the spherically-averaged power spectrum. We also add the Alcock-Paczy\'{n}ski effect \cite{AP1979, AliAP, Nusser, APeffect} and the light-cone effect \cite{barkana06, mondal18} to our power spectrum calculations, and account for the effect of the field of view and angular resolution of radio interferometers (see {\it Supplementary Information}). We use the latest measurements (based mainly on the CMB)~\cite{Planck:2018} to set our fiducial values of the cosmological parameters. The 21-cm global signal and power spectra during the dark ages for this fiducial model are shown, respectively, in Figs.~\ref{fig:global} and \ref{fig:power}, which are explained below in further detail.


\subsection*{The global 21-cm signal}
\label{sec:global}
\begin{figure}[ht]
\centering
\includegraphics[width=.9\textwidth, angle=0]{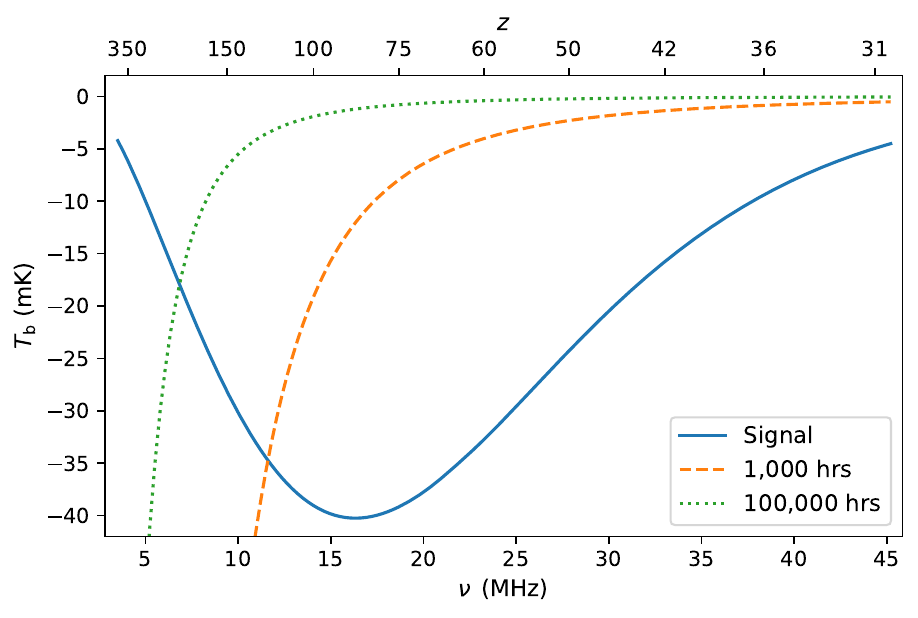}
\caption{{\bf The 21 cm global signal from the dark ages.} The 21 cm signal is shown as a function of $\nu$ (and $z$ as the top $x$ axis). We also show the expected thermal noise for a global signal experiment observing for integration times of 1,000\,hrs and 100,000\,hrs for bins with $\Delta(\ln \nu) = 1$.}
\label{fig:global}
\end{figure}

As noted above, measuring the 21-cm global signal requires a single, well-calibrated antenna. Fig.~\ref{fig:global} shows the 21-cm global signal from the dark ages as a function of $\nu$ (and $z$), for the fiducial cosmological model. The expected signal dips to a maximum absorption of $-40.2$\,mK at $z=86$ ($\nu=16.3$~MHz). Also shown in Fig.~\ref{fig:global} is the instrumental noise for $t_{\rm int}=1$,000\,hrs (a standard fiducial integration time, also equal to 11.4\% of a year) and 100,000\,hrs, for a bin of width $\Delta(\ln \nu) = 1$ around each $\nu$. The noise increases sharply with redshift, yielding a maximum signal-to-noise ratio (S/N) of 11.6 for $t_{\rm int}=1$,000\,hrs and 116 for 100,000\,hrs (both at $z=41$ or $\nu=34$~MHz); in the latter case, the S/N for the global signal remains above unity up to $z = 207$ ($\nu = 6.8$\,MHz). We note that if other difficulties are overcome and integration time becomes the main issue, then multiple copies of a global experiment effectively increase $t_{\rm int}$ in proportion to the number of copies (as long as they are not placed spatially too close together).

By fitting the standard cosmological model to the global signal with its expected errors, we find the constraints that can be obtained on the cosmological parameters. While the amplitude of the global signal depends significantly on some of the parameters, the shape is highly insensitive, which means that the signal essentially measures a single quantity, the overall amplitude, which depends on a combination of cosmological parameters. Specifically, the global signal depends significantly only on the parameters $\Omega_{\rm b}h^2$ and $\Omega_{\rm m}h^2$, where $\Omega_{\rm b}$ and $\Omega_{\rm m}$ are the cosmic mean densities of baryons and (total) matter, respectively, in units of the critical density, and $h$ is the Hubble constant in units of 100~km~s$^{-1}$~Mpc$^{-1}$. Given the strong degeneracy between the two parameters, the constraint is on the combination (see {\it Supplementary Information})
\begin{equation}
\label{eq:CGlobal}
C_{\rm Global} \equiv \frac{\Omega_{\rm b}h^2}{(\Omega_{\rm m}h^2)^{0.248}}\ .
\end{equation}

The relative errors in $C_{\rm Global}$ for three different values of $t_{\rm int}$ are shown (along with our other main results) in Table~\ref{tab:error} and Fig.~\ref{fig:relative_error}. We account for the fact that the presence of the bright synchrotron foreground means that a signal component of the same shape cannot be distinguished from the foreground (see {\it Supplementary Information}). A measurement of the global 21-cm signal to the precision of thermal noise from a 1,000~hour integration would yield a 10.1\% measurement. This would be a remarkable achievement for cosmological concordance, since it would be independent of other cosmological probes and come from a previously unexplored cosmological era. Increasing the integration time would improve this precision as the inverse square root, so that sub-percent precision in $C_{\rm Global}$ (comparable to the typical Planck precision on each cosmological parameter) would require a considerable integration time exceeding 100,000~hrs.


\begin{table}[ht]
\begin{center}
\caption{The relative errors in \% and the limits on the total mass of neutrinos (all are $1\sigma$). For the Helium fraction ($Y_{\rm P}$) and the neutrino mass, we compare to constraints based on Planck CMB measurements alone and also (Planck + BAO) those that include BAO measurements from galaxy clustering~\cite{Planck:2018}.}
\label{tab:error}

\begin{minipage}{9.5cm}
\begin{tabular*}
{\textwidth}{@{\extracolsep{\fill}}lccccc@{\extracolsep{\fill}}}
\toprule%

{\large Global} & {Planck} & {Planck} & \multicolumn{3}{@{}c@{}}{\underline{~~~~~~~~~~~~~~Integration time~~~~~~~~~~~~~~}}\\


{\large signal} & {$+$\,BAO} &  & 100,000\,hrs & 10,000\,hrs & 1,000\,hrs \\

\midrule
$C_{\rm Global}$ & & & 1.01 & 3.18 & 10.1 \\

$Y_{\rm P}$ & 4.96 & 5.44 & 3.14 & 9.94 & 31.4 \\

$\sum m_{\nu}\,[{\rm eV}]$ & $<0.0578$ & $<0.108$ & $< 0.746$ \\
\vspace{-0.2cm}
\end{tabular*}
\end{minipage}

\begin{minipage}{\textwidth}
\begin{tabular*}
{\textwidth}{@{\extracolsep{\fill}}lccccccc@{\extracolsep{\fill}}}
\toprule

{\large Power} & {Planck} & {Planck} &  \multicolumn{5}{@{}c@{}}{\underline{~~~~~~~~~~~~~~~~~~~~~~~~~~~Configuration~~~~~~~~~~~~~~~~~~~~~~~~~}}\\

{\large spectrum} & {$+$\,BAO} & & D & C & B & A & G \\

\midrule

$C_{\rm PowSpec}$ &  &  & 0.0457 & 0.382 & 0.462 & 4.59 & 10.1 \\

$Y_{\rm P}$ & 4.96 & 5.44 & 0.116 & 0.981 & 1.20 & 11.9 & 26.6 \\

$\sum m_{\nu}\,[{\rm eV}]$ & $<0.0578$ & $<0.108$ & $<0.0100$ & $<0.0839$ & $< 0.107$ & $< 1.06$ \\

\botrule
\end{tabular*}

\end{minipage}
\end{center}
\end{table}


In addition to the minimal, standard set of cosmological parameters, the global signal can also provide additional constraints, of which we consider a couple examples. For these additional parameters, we consider the favorable approach in which we fix the standard set of parameters at their fiducial values (e.g., based on Planck with its small errors) and explore the power of the 21-cm signal to constrain these specific extended parameters. In particular, since the 21-cm signal depends directly on hydrogen and not just the total baryon density, the first additional parameter is the fraction of the baryonic mass in helium, usually denoted $Y_{\rm P}$. This parameter is currently constrained by Planck (even when combined with galaxy clustering) at a level that is an order of magnitude worse than the precision on the standard parameters. For $t_{\rm int}=1$,000\,hrs, the global 21-cm signal would yield an independent 31.4\% constraint on $Y_{\rm P}$; 10,000\,hrs would measure $Y_{\rm P}$ to 9.94\%, and the best-case scenario of 100,000\,hrs would beat Planck by a factor of 1.7 (Table~\ref{tab:error} and Fig.~\ref{fig:relative_error}). Similarly, we find constraints on the neutrino mass, though for this purpose the global signal would not be competitive with Planck, even for $t_{\rm int}=100$,000\,hrs.


\subsection*{The 21-cm power spectrum}
\label{sec:power}
\begin{figure}
\centering
\includegraphics[width=.9\textwidth, angle=0]{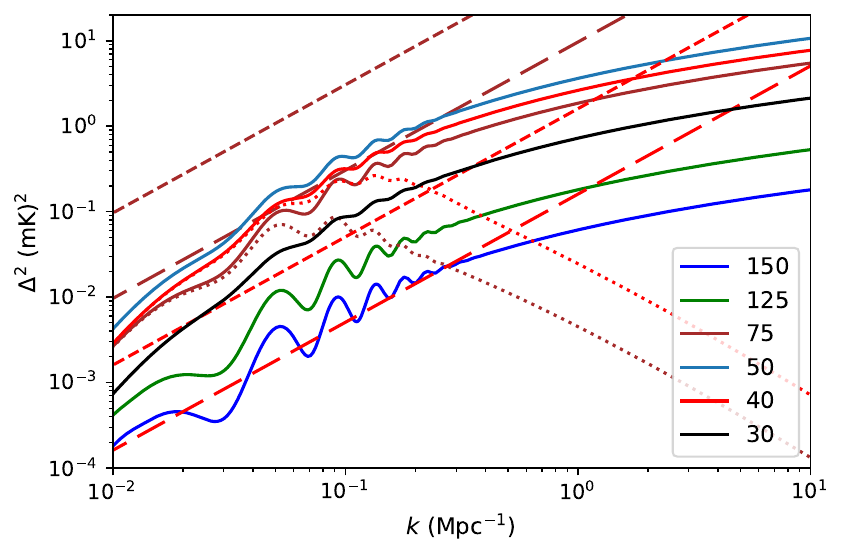}
\caption{{\bf The spherically averaged (total) power spectrum of 21 cm brightness fluctuations.} The fluctuations are shown as a function of $k$ during the dark ages at $z = [150, 125, 75, 50, 40, 30]$. The dotted lines show the power spectra at $z = 75$ and 40 when accounting for the effect of angular resolution (for our A or B configurations). We also show the 1$\sigma$ noise (thermal plus cosmic variance) for our A (short dashed lines) and B (long dashed lines) configurations at $z = 75$ and 40 (for bins with $\Delta(\ln \nu) = 1$ and $\Delta (\ln k) = 1$).}
\label{fig:power}
\end{figure}

As noted above, compared to the global 21-cm signal from the dark ages, it would take a substantially larger effort to measure the power spectrum. However, looking towards the future, the power spectrum has a far greater potential to become a ground-breaking cosmological probe, as it is a much richer dataset. Fig.~\ref{fig:power} shows the spherically-averaged power spectrum of 21-cm brightness fluctuations as a function of wavenumber at various redshifts during the dark ages. The signal rises initially as the adiabatic expansion cools the gas faster than the CMB, creating an absorption signal that strengthens with time. Eventually, though, the declining density reduces the collisional coupling so that $x_c$ drops below unity and the 21-cm signal weakens. For example, the maximum squared fluctuation $\Delta^2$ at $k=0.1\,{\rm Mpc}^{-1}$ is 0.44~mK$^2$ at $z=51$. 

For the observational setup, we assume a minimal case for which a 1,000 hour integration would significantly exceed (by a factor of $\sim$ 2) the constraint level given by the same global case; this would be required to justify the much greater effort involved in building an interferometer. We find that this would (approximately) require a collecting area of $A_{\rm coll} = 10\,{\rm km}^2$, which along with $t_{\rm int}=1$,000\,hrs we adopt as our minimal, A configuration. The collecting area of 10\,km$^2$ corresponds to 400,000 stations, each with an effective collecting area of 25\,m$^2$ (see {\it Supplementary Information}). This is quite futuristic but we hope that our theoretical work helps motivate new ideas to achieve this more quickly. The four observational configurations that we use to illustrate measurements of the 21-cm power spectrum are listed in Table~\ref{tab:conf} (for reference, we also include a G configuration that yields constraints roughly equal to the 1,000 hour global case). Fig.~\ref{fig:power} shows the 1$\sigma$ noise expected for our A and B configurations, when we include the (dominant) thermal noise as well as cosmic variance (see {\it Supplementary Information}). The figure also shows the power spectrum when accounting for the effect of angular resolution. The thermal noise increases rapidly with redshift, and so the maximum S/N (without the effect of angular resolution) occurs at the minimum redshift we consider ($z=30$), and is 13.3 for the A configuration and 133 for the B configuration, both at $k=0.091$\,Mpc$^{-1}$. 


\begin{table}[ht]
\begin{center}
\caption{{\bf The observational configurations used to illustrate measurements of the 21 cm power spectrum} (see {\it Supplementary Information}).}
\label{tab:conf}
\begin{minipage}{8.5cm}
\begin{tabular*}
{\textwidth}{@{\extracolsep{\fill}}lccccc@{\extracolsep{\fill}}}
\toprule%

& \multicolumn{5}{@{}c@{}}{Configuration}\\
\cmidrule{2-6}
{} & D & C & B & A & G\\

\midrule

$A_{\rm coll}$ [km$^2$]& 100 & 100 & 10 & 10 & 5 \\
$t_{\rm int}$ [hrs]& 10,000 & 1,000 & 10,000 & 1,000 & 1,000 \\

\botrule
\end{tabular*}
\end{minipage}
\end{center}
\end{table}


\begin{figure}
\centering
\includegraphics[width=.9\textwidth, angle=0]{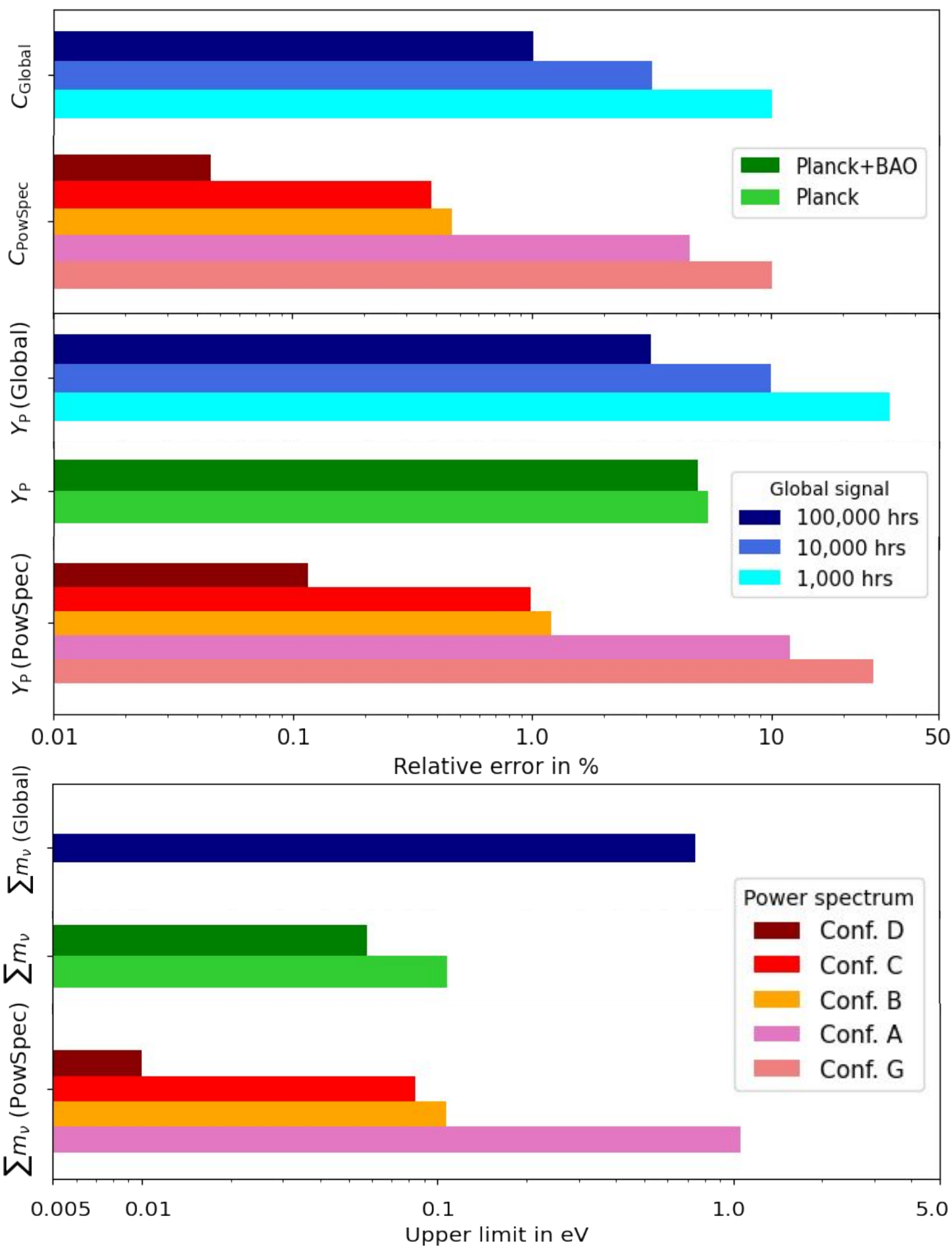}
\caption{{\bf The relative errors and limits on the total mass of neutrinos.} We show the main results that are listed in Table~\ref{tab:error}, namely the relative errors (top) and the limits on the total mass of neutrinos (bottom) (all are 1$\sigma$). Conf., configuration.}
\label{fig:relative_error}
\end{figure}

Given measurements of the 21-cm power spectrum throughout the dark ages ($z>30$), we carry out a Fisher analysis with five cosmological parameters (see {\it Supplementary Information}). The relative errors in the $\Lambda$CDM cosmological parameters are rather large due to significant degeneracies, and even configuration D approaches the accuracy level of Planck only in some of the parameters (see {\it Supplementary Information}). As with the global signal, it is more useful to consider parameter combinations that are well constrained. In particular, we focus on the minimum variance combination (see {\it Supplementary Information}), which for configuration A is
\begin{equation}
C_{\rm PowSpec} \equiv \Omega_{\rm b}h^2 \frac{ (A_{\rm s}e^{-2\tau})^{0.307} (0.9950)^{n_{\rm s}}}{(\Omega_{\rm m}h^2)^{0.464} H_0^{0.0753}} \, .
\label{eq:cpower}
\end{equation}
Here the additional parameters~\cite{Planck:2018} are the Hubble constant $H_0$ (in units of km\,s$^{-1}$\,Mpc$^{-1}$), the primordial amplitude $A_{\rm s}$, the total reionization optical depth to the CMB $\tau$, and the scalar spectral index $n_{\rm s}$. We note that the form of $C_{\rm PowSpec}$ (eq.~\ref{eq:cpower}) changes slightly for different scenarios (see {\it Supplementary Information}). 

The relative errors in $C_{\rm PowSpec}$ for the various observational configurations are shown in Table~\ref{tab:error} and Fig.~\ref{fig:relative_error}. Configuration A would yield a 4.59\% measurement of the parameter combination $C_{\rm PowSpec}$, which would be observationally independent of the global signal constraint and thus provide a powerful cross-check. More importantly, there would be a great potential for future improvement, as Configurations B and (the slightly better) C would improve this by an order of magnitude (reaching the typical Planck precision on each cosmological parameter), and D by a further order of magnitude. Just as for the global signal, we consider constraints on additional parameters while fixing the standard set of parameters. Here configuration A would measure $Y_{\rm P}$ to 11.9\%, B and C would do 5 times better than the Planck constraint, and D would do almost 50 times better than Planck. It is reasonable to again consider these constraints while fixing the standard parameters based on Planck, since Planck constrains the standard parameters better than any of our 21-cm configurations. Finally, the constraint on the total neutrino mass would not be competitive with Planck for configuration A, but would roughly match Planck for configurations B or C, and beat it by an order of magnitude for configuration D. This constraint is driven by the suppression of small-scale power due to neutrino free-streaming. Here we emphasize a major advantage for probing the neutrino effect on gravitational clustering during the dark ages: the corresponding scales were still in the regime of linear fluctuations, and were not yet affected by the complex astrophysics of galaxies. 

\section*{Conclusions}
\label{sec:conc}
Observations of the redshifted 21-cm signal from the dark ages have great cosmological potential. While various exotic, non-standard scenarios could be easily detected (or ruled out), here we considered the safe, conservative case of standard cosmology, studying the potential for creating a powerful new cosmological probe. We found constraints on the $\Lambda$CDM cosmological parameters, independently considering the two major types of 21-cm measurements, the global (or mean) signal as a function of frequency and the spherically-averaged power spectrum as a function of both frequency and scale. We used \texttt{CAMB} and added to it redshift space distortions, the Alcock-Paczy\'{n}ski effect, the light-cone effect, and the effect of angular resolution. For the error estimates, we considered different levels of thermal noise (plus cosmic variance), meant to serve as a benchmark for experiments which face additional practical challenges including foreground removal.

With global 21-cm signal measurements, we found that a combination of cosmological parameters, $C_{\rm Global}$ (eq.~\ref{eq:CGlobal}), can be effectively constrained. An integration time of 1,000\,hrs would yield a relative error in $C_{\rm Global}$ of 10.1\%, with improvement to a best-case precision of 1.01\% for 100,000\,hrs. In the case of the 21-cm power spectrum, it would take a greater effort to achieve comparable constraints, but there are better prospects for future advances. The parameter combination $C_{\rm PowSpec}$ (eq.~\ref{eq:cpower}) can be constrained to 4.59\% in our configuration A (a 1,000\,hr integration with an array of collecting area 10~km$^2$), but the precision can improve to 0.0457\% in our configuration D (a 10,000\,hr integration with a collecting area of 100~km$^2$). 

Fixing the standard set of cosmological parameters to their fiducial values, we found constraints on separately varying two other important parameters. Given the direct dependence of the 21-cm signal on hydrogen, the fraction of the baryonic mass in helium $Y_{\rm P}$ would be constrained to 31.4\% with a 1,000\,hr integration of the global signal; 10,000\,hrs would measure it to 9.94\%,
and the best-case scenario of 100,000\,hrs would beat Planck by a factor of 1.7. Using the power spectrum, configuration A would measure $Y_{\rm P}$ to 11.9\%, B and C would do 5 times better than the Planck constraint, and D would do almost 50 times better than Planck. Regarding limits on the total mass of neutrinos, constraints that are competitive with Planck would be possible only with the 21-cm power spectrum, for which configurations B or C would roughly match Planck, and configuration D would beat it by an order of magnitude.

Our analysis highlights the potential of the 21-cm signal as a probe of cosmology, and suggests a focus on the global signal as the first step, with the 21-cm power spectrum being much more promising in the long run. Our results set a baseline reference for many upcoming and future lunar and space-based dark ages experiments.




\section*{Data availability}
The data are available upon request from the corresponding author. Source data are provided with this paper.

\section*{Code availability}
\texttt{CAMB} is available at \url{http://camb.info}. \texttt{emcee} is available at \url{https://github.com/dfm/emcee}. \texttt{corner} is available at \url{https://github.com/dfm/corner.py}. The analyses are done in Python using publicly available routines in NumPy (\url{https://numpy.org}) and Matplotlib (\url{https://matplotlib.org}). All other codes used are available upon request from the corresponding author.

\section*{Acknowledgments}
We thank A. Lewis and L. V. E. Koopmans for their useful discussions. R.M. is supported by the Israel Academy of Sciences and Humanities \& Council for Higher Education Excellence Fellowship Program for International Postdoctoral Researchers. We acknowledge the support of the Israel Science Foundation (grant no. 2359/20).

\section*{Author contributions}
R.B. initiated the project. R.M. performed the calculations, made the figures, and wrote the paper, in consultation with R.B..

\section*{Competing interests}
The authors declare no competing interests. \\

\noindent {\bf Correspondence} should be addressed to Rajesh Mondal.




\begin{appendices}

\section{Supplementary note}
\label{sec:methods}
In this Supplementary Note, we first (Sec.~\ref{sec:summary}) briefly summarize our methods and add some details and technical notes. We next describe in detail how our predicted signal accounts for several effects: the Alcock-Paczy\'{n}ski effect (Sec.~\ref{sec:AP_effect}), the light-cone effect (Sec.~\ref{sec:LC_effect}), and the effect of angular resolution (Sec.~\ref{sec:resolution}). We then note our method for constructing a useful (minimum variance) linear combination of correlated parameters (Sec.~\ref{sec:degen}), and present some additional results and discussion (Sec.~\ref{sec:misc}). Finally, we briefly discuss the effect of foregrounds (Sec.~\ref{sec:foregrounds}).

\subsection{Summary of our methods}
\label{sec:summary}
The main quantity for 21-cm observations, the excess brightness temperature relative to the CMB from redshift $z$, is 
\begin{equation}
T_{\rm b} = (T_{\rm s} - T_{\gamma})\frac{1 - e^{-\tau_{21}}}{1 + z} \ ,
\end{equation}
where $\tau_{21}$ is the optical depth of the 21-cm transition. Assuming $\tau_{21} \ll 1$, this can be expressed in the simpler form  \cite{madau97, Furlanetto2006}
\begin{equation}
{T}_{\rm b} \simeq 54.0\,{\rm mK}\, \frac{\rho_{\rm HI}}{\bar{\rho}_{\rm H}} \left(\frac{\Omega_{\rm b}h^2}{0.02242}\right) \left(\frac{\Omega_{\rm m}h^2}{0.1424}\right)^{-\frac{1}{2}} \left(\frac{1 + z}{40}\right)^{\frac{1}{2}} \frac{x_{\rm c}}{1 + x_{\rm c}} \left(1 - \frac{T_{\gamma}}{T_{\rm g}}\right) ,
\label{eq:Tb}
\end{equation}
where $\rho_{\rm HI}$ is the neutral hydrogen density and $\bar{\rho}_{\rm H}$ is the cosmic mean density of hydrogen, and also $x_{\rm c}$ is the collisional coupling coefficient. During the dark ages, CMB scattering pulls $T_{\rm s} \xrightarrow[]{} T_{\gamma}$, whereas atomic collisions pull $T_{\rm s} \xrightarrow[]{} T_{\rm g}$.

As noted in the main text, we use the standard \texttt{CAMB} ({\url{http://camb.info}}) \cite{Lewis2007,CAMB} cosmological perturbation code to generate the predicted 21-cm signal. Note that \texttt{CAMB} does not directly yield the 21-cm global signal (i.e., the mean 21-cm brightness temperature) as a function of redshift, so we extract it indirectly by running once with temperature (mK) units on and once with temperature units off, and taking the ratio of the transfer functions in the two cases. Also, \texttt{CAMB} outputs the 2D angular power spectrum, inspired by CMB analyses but less relevant to 21-cm data that naturally constitute a 3D dataset both theoretically and observationally. \texttt{CAMB} yields the transfer function for the 21-cm monopole power spectrum, to which we add by hand the redshift space distortions using the transfer function of baryon density, based on Ref.~\cite{BLlos}. Having obtained the anisotropic 3D power spectrum, we then average over angle in this linear-theory case (this is shown explicitly within the derivation in Sec.~\ref{sec:AP_effect} below). After this, we add several effects that are presented in detail in the next few sections. 

In \texttt{CAMB} and throughout the paper, for the cosmological parameters we use fiducial values (based mainly on the CMB)~\cite{Planck:2018} of $H_0 =67.66$~km~s$^{-1}$~Mpc$^{-1}$, $\Omega_{\rm b}h^2 = 0.02242$, $\Omega_{\rm m}h^2 = 0.14240$, $A_{\rm s}e^{-2\tau} = 1.881\times10^{-9}$, and $n_{\rm s}= 0.9665$. We note that $\Omega_{\rm m} = \Omega_{\rm c} + \Omega_{\rm b} + \Omega_{\nu}$, where $\Omega_{\rm c} h^2 = 0.11933$ is the contribution of cold dark matter and the fiducial neutrino contribution (based on the minimal total mass of 0.06~eV allowed by neutrino oscillation experiments) is $\Omega_{\nu} h^2 = 6.451\times 10^{-4}$.

In this paper, our variables in the global signal case are $\log(\Omega_{\rm b} h^2)$ and $\log(\Omega_{\rm m} h^2)$. For the 21-cm power spectrum we have three additional variables: $\log(A_{\rm s}e^{-2\tau})$, $n_{\rm s}$, and $\log(H_0)$. We assume a flat Universe, where the rest of the energy density ($1-\Omega_{\rm m}$) is given by a cosmological constant. We note that in the 21-cm power spectrum from the dark ages (which, like cosmic recombination in the case of the CMB, occurred long before any significant reionization), the amplitude that is directly probed is $(A_{\rm s}e^{-2\tau})$ and not $A_{\rm s}$. This is since the re-scattering of the 21-cm photons during reionization damps the fluctuations as in the case of sub-horizon CMB fluctuations, i.e., the brightness temperature relative to the mean gets multiplied by a factor of $e^{-\tau}$. Unlike the CMB, there is no separate information on $A_{\rm s}$ and $\tau$, since in the CMB there is information on the largest scales and on polarization, but both of these are not expected in 21-cm measurements (at least in the near future). We also note that there is no logarithm on $n_{\rm s}$ since it is a power, i.e., it effectively is already defined as a logarithmic variable. For the power spectrum $P(k)$, we express the results 
in terms of the squared fluctuation $\Delta^2 \equiv k^3 P(k)/(2 \pi^2)$.

The thermal noise in a global signal measurement is \cite{Shaver1999}
\begin{equation}
\Delta T = \frac{T_{\rm sys}}{\sqrt{\Delta \nu \, t_{\rm int}}} \ ,
\label{eq:thermal_global}
\end{equation}
where $\Delta \nu$ is the bandwidth, $t_{\rm int}$ is the integration time, and we assume that the system temperature $T_{\rm sys}$ is approximately equal to the sky brightness temperature $T_{\rm sky} = 180\times (\nu/180\,{\rm MHz})^{-2.6}$\,K \cite{Furlanetto2006}.

Next we estimate the observational errors for the 21-cm power spectrum. Although it is negligible in most practical cases, for completeness (and for comparison with previous theoretically-motivated work) we include the error due to cosmic variance\,(CV), which for the power spectrum measured in a bin centered at a wavenumber $k$ and frequency $\nu$ we can express as (following eq.~30 of Ref.~\cite{mondal16})
\begin{equation}
\delta P_{\rm cv}(k, \nu) = \frac{2\pi P(k, \nu)}{\sqrt{V(\nu) k^3 \Delta(\ln k)}} \ , 
\end{equation}
where the survey volume for the frequency (redshift) bin $V(\nu)$, in the limit of a thin bin, is given by $\Omega_{\rm FoV} r_{\nu}^2 \Delta r_{\Delta \nu}$, where $\Omega_{\rm FoV}$ is the field of view, $r_{\nu}$ the comoving distance to the bin center, and $\Delta r_{\Delta \nu}$ is the comoving length corresponding to the bandwidth $\Delta \nu$. Note that the field of view\,(FoV) is $[21(1+z)\,{\rm cm}]^2/A_{\rm eff}$ for an antenna with an effective collecting area of $A_{\rm eff}$. For example, $\Omega_{\rm FoV} = 8.89$ at $z \sim 70$ assuming $A_{\rm eff} = 25\,{\rm m}^2$, compared to the whole sky which is $4 \pi = 12.6$. Thus, given the small effective area and large $z$, for a dark ages array the FoV is typically a significant fraction of the sky; we set a cutoff of half the sky as the maximum solid angle available to an interferometer. 

The dominant error that we include in the power spectrum measurement is that due to thermal noise, which can be expressed as \cite{mellema13}
\begin{equation}
\delta P_{\rm thermal} = \frac{2}{\pi} \left(\frac{k^{3}\, V}{\Delta (\ln k)}\right)^{1/2} \, \frac{T^2_{\rm sys}}{\Delta \nu~ t_{\rm int}} \, \frac{1}{N^2} \, \frac{A_{\rm core}}{A_{\rm eff}} \, ,
\label{eq:thermal_PS}
\end{equation}
where $N$ is the total number of stations and $A_{\rm core}$ is the core area of the telescope array. A reasonable plan for an upcoming lunar array (Leon Koopmans, personal communication) consists of $N=128^2=16,384$ stations with $A_{\rm eff} = 25\,{\rm m}^2$ and a core area equal to the collecting area, i.e., $A_{\rm core}=A_{\rm coll} = N \times A_{\rm eff}$. This gives a total collecting area of 0.4096~km$^2$. We keep all of these relations fixed but modify $N$, getting a total dependence of $\delta P_{\rm thermal} \propto 1/N$. This number must be increased by a factor of 24.4 to $N = 400$,000 to give our A configuration in the main text (with a collecting area of $10\,{\rm km}^2$). We note though that the smaller collecting area would suffice to put new strict limits on various non-standard models, and it would approach the performance of our A configuration if used with an integration time of a few tens of thousands of hours.

We showed noise estimates that are independent of binning, so that we gave the overall S/N based on a bin size of order the central value, i.e., $\Delta(\ln \nu) = 1$ as well as $\Delta(\ln k) = 1$. For the Fisher matrix predictions, we used 8 frequency/redshift bins in the range $5.81 \le \nu \le 45.81$ with a bin width of $\Delta \nu = 5\,{\rm MHz}$, which corresponds to central redshifts of [170, 106, 76.6, 59.9, 49.2, 41.6, 36.1, 31.8]; the upper end of the frequency range was chosen at $z=30$, the typical redshift where galaxies at cosmic dawn form in sufficient numbers to significantly affect the 21-cm signal~\cite{subtle}. For the power spectrum, we used 11 logarithmic $k$ bins covering the range $0.00779 \le k < 1.91\,{\rm Mpc}^{-1}$ with bin width $\Delta (\ln k) = 0.5$. We checked that the results are insensitive to increasing these binning resolutions. See also sec.~\ref{sec:misc} where these ranges are varied.  

\subsection{The Alcock-Paczy\'{n}ski effect}
\label{sec:AP_effect}
When using the 21-cm power spectrum for constraining cosmological parameters, it is important to account for the fact that the 3D power spectrum depends on distances, but these are usually not directly measurable in cosmology. Instead, redshifts are measured along the line of sight, while angles are measured on the sky. The conversions of these quantities to comoving distances depend on the values of the cosmological parameters, which themselves are being constrained by the data. This leads to the so-called Alcock-Paczy\'{n}ski effect~\cite{AP1979}, which is important also for the 21-cm signal \cite{AliAP,Nusser,APeffect}.

Following the analysis of this effect on the 21-cm power spectrum in Ref.~\cite{APeffect}, the setup is that we have a true cosmology (which we take as that given by the central, fiducial values of the cosmological parameters), and a different assumed cosmology (for example, where one of the parameters is varied from its fiducial value in order to find the resulting derivative of the signal, for the Fisher matrix calculation). The conversion to distances at redshift $z$ involves (on the sky) the angular diameter distance $D_A$ and (for small distances along the line of sight) the Hubble constant $H$ at $z$. The ratio $D_A$(true)/$D_A$(assumed) we designate $1+\alpha_\perp$, and the ratio $[H D_A]$(true)/$[H D_A]$(assumed) we designate $1+\alpha$. Note that these standard scalings, as written, are for physical distances, while we are interested in comoving distances, but this does not matter here since the difference is a redshift factor which in 21-cm cosmology is known precisely, independently of the cosmological parameters. Now, instead of using the full (complicated) equations in Ref.~\cite{APeffect}, we show here how to implement the effect of the changing distance measures in two steps. Note that to linear order the effects of $\alpha_\perp$ and of $\alpha$ are independent \cite{APeffect}. 

The first step is to include the effect of $\alpha_\perp$ assuming $\alpha=0$, which corresponds to assuming that the scalings from angle to perpendicular distance and from frequency to line-of-sight distance are the same, in terms of the true parameters relative to the assumed parameters. This case is isotropic and is simple to do exactly (without a linear approximation) \cite{APeffect}. The formulas simplify further when applied to the dimensionless combination $k^3 P(k)$ (which is proportional to $\Delta^2$):
\begin{equation}
k^3 P(k) = k_{\rm{true}}^3 P_{\rm{true}}(k_{\rm{true}})\ ,
\label{eq:alpha_p}
\end{equation}
where $k_{\rm{true}}=k/(1+\alpha_\perp)$. 


\begin{figure*}
\centering
\includegraphics[width=.49\textwidth, angle=0]{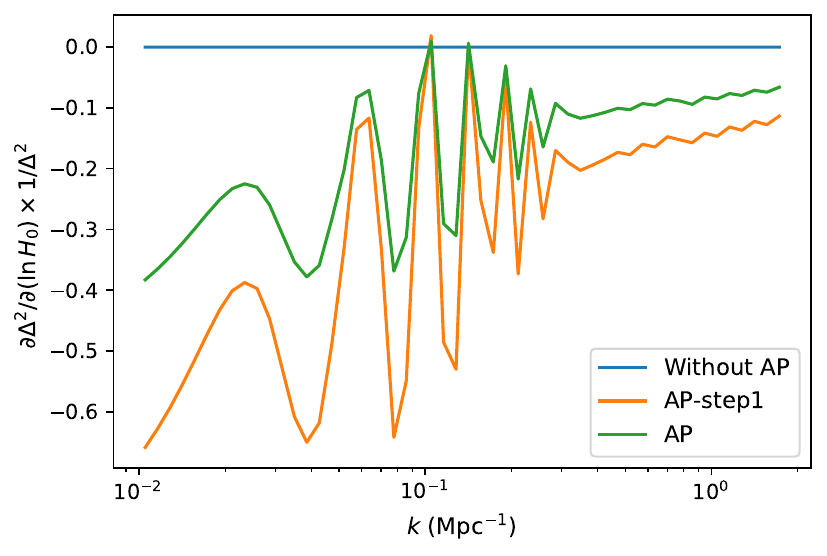}
\includegraphics[width=.48\textwidth, angle=0]{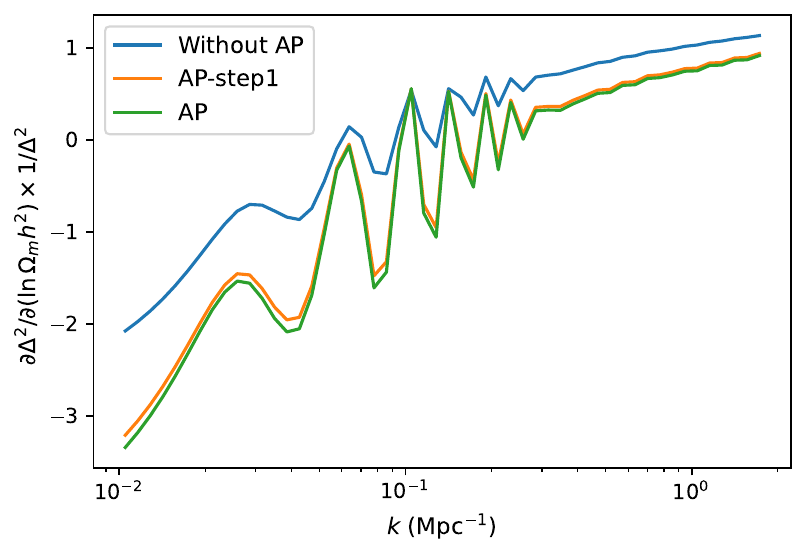}
\caption{The logarithmic dependence of the squared fluctuation on the two parameters of relevance for the Alcock-Paczy\'{n}ski (AP) effect; i.e., $\frac{1}{\Delta^2} \frac{\partial \Delta^2}{\partial(\ln H_0)}$ (left panel) and $\frac{1}{\Delta^2} \frac{\partial \Delta^2}{\partial(\ln [\Omega_{\rm m}h^2])}$ (right panel), shown at $z=40$. We show the results in three cases: without the AP effect, with the first (isotropic $\alpha_\perp$) step only, and with the full AP effect.}
\label{fig:AP}
\end{figure*}


The second effect, that of $\alpha \ne 0$, is anisotropic, but here we insert it only for the case of interest, i.e., the simplified case of the effect on the spherically-averaged power spectrum, to linear order in changes of the parameters (i.e., to first order in $\alpha$). The result uses the angular decomposition of the 21-cm power spectrum in linear theory, including  the effect of line-of-sight velocity gradients \cite{BLlos}:
\begin{equation}
P(k,\mu) = \mu^4
P_{\mu^4}(k) + \mu^2 P_{\mu^2}(k) + P_{\mu^0}(k) \ ,
\end{equation}
where $\mu = k_z/k$ is the cosine of the angle between the $\vec{k}$ vector and the line of sight. Here and subsequently, the  components (in the decomposition in powers of $\mu$) of $P(k,\mu)$ refer to the result of eq.~\ref{eq:alpha_p} (note that $\mu$ is left unchanged by the $\alpha_\perp$ rescaling). We note that $P_{\mu^0}(k)$ is the (monopole) 21-cm power spectrum without velocity effects, $P_{\mu^4}(k)$ is simply the power spectrum of the baryon density (the dimensionless power spectrum times the global temperature squared, to get mK$^2$ units), and $P_{\mu^2}(k)$ is the cross term of the 21-cm fluctuation and the baryon density fluctuation with an added factor of 2. The total spherically-averaged 21-cm power spectrum is then:
\begin{equation}
P(k) = \frac{1}{5} P_{\mu^4}(k) + \frac{1}{3} P_{\mu^2}(k) + P_{\mu^0}(k) \ .
\end{equation}
Now, from Ref.~\cite{APeffect} we find that the second effect, that of $\alpha \ne 0$, on the spherically-averaged 21-cm power spectrum, is the addition to $k^3 P(k)$ of:
\begin{equation}
\alpha\, \frac{\partial}{\partial \log k} 
\left[ \frac{1}{7}\, k^3 P_{\mu^4}(k) + \frac{1}{5}\, k^3 P_{\mu^2}(k) + \frac{1}{3}\, k^3 P_{\mu^0}(k) \right] \ ,
\end{equation}
where again the use of the dimensionless combination $k^3 P(k)$ simplified the result. Note that here the $\partial/\partial \log k$ refers to a derivative at fixed cosmological parameters (as the change in the parameters is captured through $\alpha_\perp$ and $\alpha$).

As noted in the main text, the Alcock-Paczy\'{n}ski effect is important in fitting the dark ages 21-cm power spectrum, since it introduces a dependence on $H_0$ that is separate from the dependence on the other cosmological parameters; it also modifies the dependence of the 21-cm signal on $\Omega_{\rm m} h^2$. Fig.~\ref{fig:AP} illustrates the logarithmic dependence of the power spectrum on these two parameters that are relevant for the Alcock-Paczy\'{n}ski effect. Both steps (in the above two-step procedure) have a comparable contribution to the $H_0$ dependence (which would otherwise be completely absent), while mainly the first (isotropic $\alpha_\perp$) step significantly enhances the dependence on $\Omega_{\rm m} h^2$.

\subsection{The light-cone effect}
\label{sec:LC_effect}

In measurements of the 21-cm power spectrum, since different positions along the line of sight correspond to different redshifts (i.e., what is observed are points along our past light cone), this results in anisotropy in the 21-cm power spectrum~\cite{barkana06, mondal18}. Here we are interested only in the spherically-averaged power spectrum, as averaged over the redshift span of each frequency bin, and the effect is then simply to average the signal over this redshift range.

To understand how this averaging works, it is easier to consider the correlation function, which is of course closely related to the power spectrum. For the correlation function at some (comoving) distance $r$, what we do is average over all pairs separated by $r$ in the observed volume. Let us call the line-of-sight comoving distance $x$ in this subsection (to avoid confusion with the redshift $z$). In the pair, let us call the two points \#1 and \#2; for each point \#1, we average over points \#2 in a spherical shell at a distance $r$ from point \#1. Actually, the shell is partially cut off at the near and far edges of the radial bin, but this can be neglected as long as the bin is large compared to the scales $1/k$ that we are interested in. Each spherical shell is symmetric about point \#1, so there is a cancellation as long as we can treat the power spectrum as a linear function of $x$, over distances of order $1/k$. We indeed assume this linear case, consistently with our overall approach.

What remains is the averaging over points \#1, so the result is simply an average of the power spectrum over comoving volume:
\begin{equation}
\frac{\int x^2 \Omega(x) P(k,x) dx}{\int x^2 \Omega(x) dx}\ .
\end{equation}
Here the solid angle $\Omega(z)$ at each $z(x)$ is the same as before (a function of $z$ but no more than half the sky). Also, $P(k,x)$ denotes the power spectrum at $k$ at the redshift corresponding to comoving distance $x$ (note that $x = (1+z) D_A(z)$ in terms of the angular diameter distance). For each frequency bin, the integrals are over the range of $x$ corresponding to the bin.

We find that the light-cone effect in our analysis is fairly small given our bandwidth of $\Delta \nu = 5$\,MHz. E.g., when fitting the power spectrum with configuration A, if we do not include the light-cone effect, the error in $C_{\rm PowSpec}$ changes from 4.59\% to 4.95\%.

\subsection{Angular resolution}
\label{sec:resolution}
When radio interferometry is done at increasingly high redshifts, it becomes more difficult to achieve a given angular resolution. Thus, the resolution is a significant limiting factor in measuring the 21-cm fluctuations from the dark ages. We account for the effect of angular resolution analytically, as follows. Based on simulations of future radio arrays as well as experience with current arrays (Koopmans, personal communication), it is a good approximation to assume a Gaussian point-spread function (PSF), with a full-width at half max (FWHM) corresponding to $0.6 \lambda /D$, where $\lambda$ is the wavelength and $D$ is the maximum diameter of the array (which we find assuming that the collecting area is a full circle). 

Thus, if the comoving coordinates are $X$, $Y$, and $Z$ (with the latter being the line-of-sight direction in this subsection), angular resolution smooths the 21-cm map with a window function
\begin{equation}
W = \frac{1} {2 \pi R^2}\, e^{-(X^2 + Y^2)/(2 R^2)} \delta_D(Z)\ ,
\end{equation}
where $\delta_D$ is a Dirac delta function, the pre-factor ensures normalization to a volume integral of unity, and $R$ is the comoving distance corresponding to angle $\theta_D$ (i.e., $R=(1+z) D_A(z) \theta_D$ in terms of the angular diameter distance $D_A$), where the above yields an angle
\begin{equation}
\theta_D = \frac{0.6 \lambda /D}{2 \sqrt{2 \ln{2}} } = 0.25\, \frac{\lambda}{D} = 
9.\mkern-4mu^\prime 1 \left( \frac{1+z}{50} \right) \left(
\frac{D}{1\,\mathrm{km}} \right)^{-1} \ .
\end{equation}
Then the Fourier transform of $W$ is 
\begin{equation}
\tilde{W} = \int d^3r\,W e^{-i \vec{k} \cdot \vec{r}} = e^{-\frac{1}{2} k^2 R^2 (1 - \mu^2)} \ ,
\end{equation}
where $\mu = \cos{\theta}$ in terms of the angle $\theta$ between $\vec{k}$ and the line of sight. The power spectrum is multiplied by the square of $\tilde{W}$. 

\begin{figure}
\centering
\includegraphics[width=.9\textwidth, angle=0]{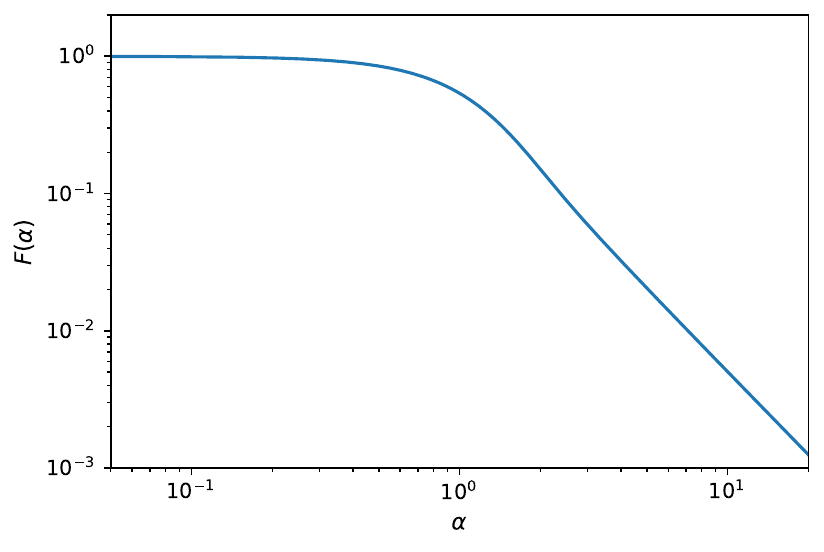}
\caption{The function $F(\alpha)$ that captures the effect of angular resolution, is shown as a function of $\alpha$.}
\label{fig:FkR}
\end{figure}

Finally, since here we are only considering the spherically-averaged power spectrum, 
we average over angle, which multiplies the power spectrum by the factor $F(kR$), where
\begin{equation}
F(\alpha) = \frac{1}{2} \int_{-1}^{1}  d \mu\, e^{\alpha^2 (\mu^2 - 1)} \ .
\end{equation}
This integral is related to the error function, but note that the coefficient of $\mu^2$ in the exponent is positive. This function is shown in Fig.~\ref{fig:FkR}.

As an example, when fitting the power spectrum with configuration A, if we do not include the effect of angular resolution, the error in $C_{\rm PowSpec}$ changes from 4.59\% to 3.53\%. We note, however, that since in our calculations, we compare the theoretically-predicted power spectrum with the effect of angular resolution to the standard expression for the thermal noise, we are being somewhat overly conservative since in reality the limited angular resolution will also smooth out the power spectrum of the thermal noise (an effect that we do not include). 

\subsection{Method for constructing a minimum variance linear combination of correlated parameters}
\label{sec:degen}
Given that the fitting of the 21-cm signal to cosmological parameters results in significant degeneracies among the parameters, we found it useful to construct combinations of the parameters that have a minimum variance. This best captures the constraining power of the data, especially since these combinations are unique to the 21-cm signal and are substantially different from the combinations that are best constrained by other cosmological datasets.

Fitting the 21-cm global signal is a case of two parameters. In general, let the parameters be $x$ and $y$, and assume we know $\sigma_x \equiv \sqrt{\langle(\Delta x)^2 \rangle}$ (where $\Delta x \equiv x - \langle x \rangle$),
$\sigma_y \equiv \sqrt{\langle(\Delta y)^2 \rangle}$, and 
the correlation coefficient $r= \langle \Delta x \Delta y \rangle /(\sigma_x \sigma_y)$. We treat $x$ as the primary variable, which in practice should be chosen as the parameter that the signal is most sensitive to; naturally, this parameter will have the largest coefficient in the linear combination below.
Then the linear combination of $x$ and $y$ with minimum variance, normalized so that $x$ has a coefficient of unity, is:
\begin{equation}
C = x - \alpha y\ ,
\end{equation}
where 
\begin{equation}
\alpha = r \frac{\sigma_x} {\sigma_y}\ ,
\end{equation}
 and $C$ has a standard deviation of
 \begin{equation}
\sigma_C = \sigma_x \sqrt{1 - r^2}\ .
\end{equation}

Fitting the 21-cm power spectrum is a case of $n=5$ parameters. In general, let the parameters be $x_i$, with $i=1$ through $n$. Then we desire to find the weights $w_i$ so that the parameter combination 
\begin{equation}
C=\sum_i x_i w_i \, ,
\end{equation}
has minimum variance, where in the weight vector $w$, we fix $w_1=1$, thus treating $x_1$ as the primary variable. Assume we know the covariance matrix $S_{ij}=\langle \Delta x_i \Delta x_j \rangle$. Then to get the solution, we remove the
first row and column and obtain the reduced $(n-1) \times (n-1)$ matrix $U$, which is simply $S_{ij}$ for $i,j>1$. Also the covariances of the other $x_j$ (for $j>1$) with $x_1$, i.e., $S_{j1}$ for $j>1$, we will
call the $(n-1) \times 1$ vector $v$. In addition, the $n-1$ weights, $w_j$
for $j>1$, are a reduced weight vector $z$. Now we solve: $U z = - v$, so that
the solution is: 
\begin{equation}
z = - U^{-1} v\ . 
\end{equation}
We construct the full vector $w$ from this solution for $z$, and the resulting minimum 
variance is 
\begin{equation}
\sigma_C^2 = w^T S w = \sum_{i,j} S_{ij} w_i w_j\ , 
\label{eq:sig_comb}
\end{equation}
where $w^T$ is the transpose of $w$, and $i,j$ go from 1 to $n$. 

We note that it is a general result that if only the parameter $x_1$ is fit to the data with all the other parameters held fixed, then the resulting error $\sigma_1$ in $x_1$ is in fact equal to the just-written expression for $\sigma_C$.

\subsection{Additional results and discussion}
\label{sec:misc}
In this section we present a number of additional results and checks, along with additional discussion. We begin with the global signal. 
We focused on the parameter combination $C_{\rm Global}$, 
where the power in the denominator indicates the power-law dependence of the global signal amplitude on $\Omega_{\rm m}h^2$ relative to $\Omega_{\rm b}h^2$. It is naturally expected to be near 1/4, given eq.~(\ref{eq:Tb}) (with its terms directly suggesting a power of 1/2) plus the fact that most of the signal-to-noise comes from the relatively low redshifts where $x_c$ is significantly below 1, and this coefficient is proportional to the collision rate per atom, and thus to $\Omega_{\rm b}h^2$; this suggests a total dependence roughly proportional to $(\Omega_{\rm b}h^2)^2/(\Omega_{\rm m}h^2)^{1/2}$, and $C_{\rm Global}$ then goes as the square root of this (since we fix the dependence on the primary parameter, $\Omega_{\rm b}h^2$, to the power of unity). In Table~\ref{tab:error_global} we also show the global signal constraints on the two relevant cosmological parameters. The errors are very large, in general and also compared to the Planck measurements. There is a nearly complete degeneracy, in that the correlation coefficient between $\Omega_{\rm b}h^2$ and $\Omega_{\rm m}h^2$, for example for $t_{\rm int}=10$,000\,hrs, is 0.994. We note that for any parameter $x$, $\sigma[\ln(x)]$ equals the relative error in $x$ when $\sigma \ll 1$; this relation is only approximately true when $\sigma$ is not small, but for simplicity, we always quote $\sigma[\ln(x)]$ as the {\it relative error} in $x$.

\begin{table}[ht]
\begin{center}
\caption{For the global signal, the 1$\sigma$ relative errors (in \%) on cosmological parameters, compared to Planck.}
\label{tab:error_global}

\begin{minipage}{9cm}
\begin{tabular*}{\textwidth}{@{\extracolsep{\fill}}lccccc@{\extracolsep{\fill}}}
\toprule%

{\large Global} & {Planck} & {Planck} & \multicolumn{3}{@{}c@{}}{\underline{~~~~~~~~~~~~~~Integration time~~~~~~~~~~~~~~}}\\
{\large signal} & {$+$\,BAO} &  & 100,000\,hrs & 10,000\,hrs & 1,000\,hrs \\

\midrule
{$\Omega_{\rm b}h^2$} & 0.624 & 0.671 & 9.76 & 30.9 & 97.6 \\
{$\Omega_{\rm m}h^2$} & 0.611 & 0.769 & 39.2 & 124 & 392 \\

\botrule
\end{tabular*}

\end{minipage}
\end{center}
\end{table}

\begin{figure}[ht]
\centering
\includegraphics[width=.9\textwidth, angle=0]{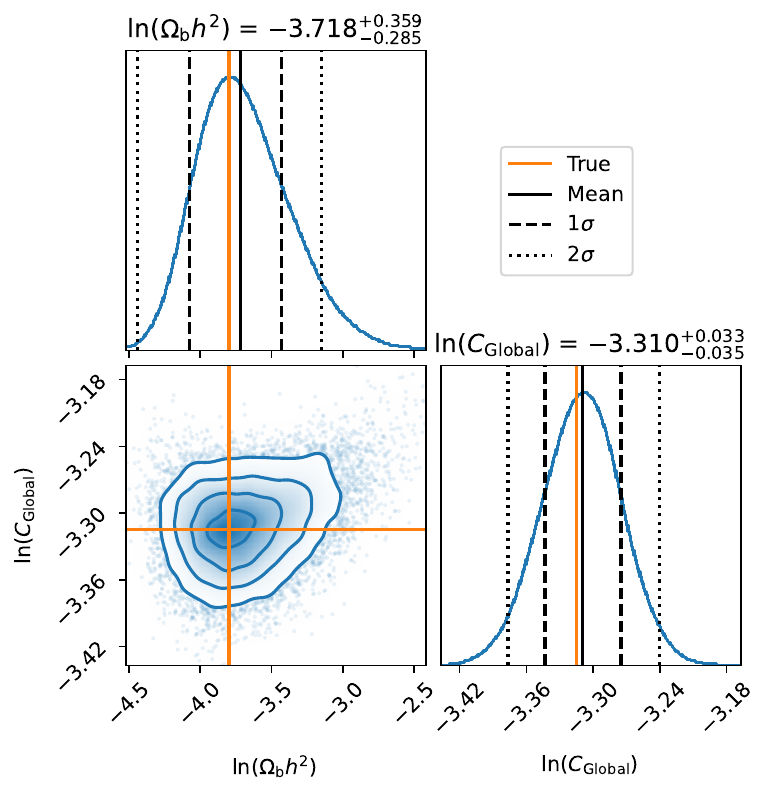}
\caption{Global 21-cm signal constraints based on MCMC fitting for $t_{\rm int} = 10$,000\,hrs. Here we use two basic variables, $\ln(\Omega_{\rm b} h^2)$ and the logarithm of $C_{\rm Global} \equiv \Omega_{\rm b}h^2/(\Omega_{\rm m}h^2)^{0.248}$. The panels show the posterior distributions (1D and 2D) of the two parameters.}
\label{fig:global_MC}
\end{figure}

As the errors on the parameters are large while that on $C_{\rm Global}$ is small, we run an MCMC chain in one case ($t_{\rm int}=10$,000\,hrs) to verify that the non-linear individual errors are not leading to a breakdown of the Fisher matrix approach as applied to the important parameter, $C_{\rm Global}$. The results are shown in Fig.~\ref{fig:global_MC}. In the 2D posterior panel, we see that there is almost no correlation between $\Omega_{\rm b}h^2$ and $C_{\rm Global}$, which justifies the choice of $C_{\rm Global}$ as the second parameter. We find a 1$\sigma$ constraint on $\ln{C_{\rm Global}}$ of $-3.310^{+0.033}_{-0.035}$, equivalent to a relative error of 3.4\% in $C_{\rm Global}$, compared to the Fisher approach that gave a relative uncertainty of $3.18\%$, an error that is close to the MCMC limits.
Also the 2$\sigma$ MCMC constraint on $\ln{C_{\rm Global}}$ is $-3.310^{+0.067}_{-0.070}$, which is roughly double the 1$\sigma$ range but shows slight asymmetry. We conclude that the Fisher approach is good enough for approximate answers in this first analysis, but full MCMC is needed for higher precision when the errors in some of the underlying parameters are large. Fig.~\ref{fig:global_MC} was generated using the python packages \texttt{emcee} \cite{Foreman-Mackey2013} and \texttt{corner} \cite{corner}.

\begin{figure}[ht]
\centering
\includegraphics[width=.9\textwidth, angle=0]{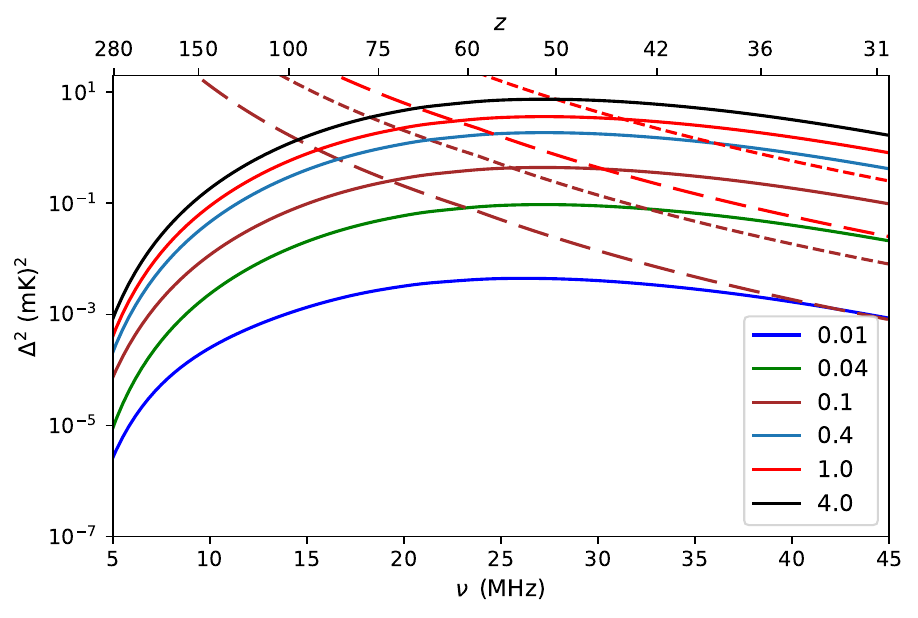}
\caption{The spherically-averaged (total) power spectrum of 21-cm brightness fluctuations as a function of $\nu$ (or $z$ as the top $x$-axis) at wavenumber values $k=[0.01$, 0.04, 0.1, 0.4, 1.0, $4.0]\,{\rm Mpc}^{-1}$. We also show the 1$\sigma$ noise (thermal plus cosmic variance) for our A (short dashed lines) and B (long dashed lines) configurations, at $k=0.1$\,Mpc$^{-1}$ and 1.0\,Mpc$^{-1}$. 
The effect of the angular resolution is not shown here.}
\label{fig:power_nu}
\end{figure}

Moving to the 21-cm power spectrum, 
in Fig.~2 in the main text we showed slices through this 2D dataset at fixed redshifts. Fig.~\ref{fig:power_nu} shows slices in the opposite direction, namely the variation with $\nu$ (or $z$), at fixed wavenumber values $k=[0.01$, 0.04, 0.1, 0.4, 1.0, $4.0]\,{\rm Mpc}^{-1}$, for the fiducial cosmological model. As expected, the power increases as we go from large scales to small scales. The power (for all the curves that are shown) peaks at $z=51$. These slices show the smooth evolution with redshift at each $k$. We also show the 1$\sigma$ noise curves (thermal plus cosmic variance) for the A and B configurations.

We now consider the results for the cosmic variance (CV) only case, which corresponds to the limit of infinite collecting area or integration time. This is a purely theoretical limit of some interest as a comparison case, given its role in some previous work \cite{Floss2022, mondal17}. We assume in this limiting case no thermal noise, perfect angular resolution, and a full sky (i.e., $\Omega_{\rm FoV} = 4 \pi$). The relative error in $C_{\rm PowSpec}$ for the CV-only case would be $7.72\times 10^{-5}\,$\%. The relative error in $Y_{\rm P}$ (fixing all other parameters) would be $2.18 \times 10^{-4}\,$\%, and the sum of the neutrino masses would be constrained to $\sum m_{\nu} < 2.40 \times 10^{-5}$\,eV. Fixing the other parameters would not be an appropriate assumption in this case with such minuscule errors, but we include this here for comparison with the other cases considered in the main text. 

As noted in the main text, there are strong correlations among the cosmological parameters, which is what led us to focus on the combination $C_{\rm PowSpec}$. The values of the correlation coefficients are illustrated in Table~\ref{tab:corr}, for Configuration A and for the CV-only case. Some of the coefficients approach unity in absolute value. 

We summarize the coefficients for configuration A 
as [0.307, 0.9950, 0.464, 0.0753] for the power of $(A_{\rm s}e^{-2\tau})$, the base of $n_{\rm s}$, and the powers in the denominator of $\Omega_{\rm m}h^2$ and $H_0$, respectively. While the dependence of the 21-cm power spectrum on the cosmological parameters is complex, we can try to roughly understand what drives the various powers in the combination $C_{\rm PowSpec}$. As discussed in the first paragraph of this section, the global signal is roughly proportional to $(\Omega_{\rm b}h^2)^2/(\Omega_{\rm m}h^2)^{1/2}$. The power spectrum goes as the global signal squared times the dimensionless (i.e., relative) squared fluctuation level. This is proportional to the primordial amplitude $A_{\rm s}$, reduced by post-reionization scattering (as for all sub-horizon scales in the CMB) by the factor $e^{-2\tau}$. Then, the growth of fluctuations (squared) from the early Universe down to the cosmic dark ages is roughly proportional to the growth factor (squared) at the dark ages relative to matter-radiation equality (which is when significant matter fluctuation growth begins). Fixing as before the dependence on the primary parameter, $\Omega_{\rm b}h^2$, to a power of unity, this would suggest a power of 0.25 for $A_{\rm s}e^{-2\tau}$ and 0.75 in the denominator for $\Omega_{\rm m}h^2$. The actual powers are changed by various additional complications, including a strong scale dependence in the sensitivity to $\Omega_{\rm m}h^2$ and a weak separate sensitivity to the Hubble constant introduced by the Alcock-Paczy\'{n}ski effect (see Sec.~\ref{sec:AP_effect}). In addition, the weak dependence on $n_{\rm s}$ in $C_{\rm PowSpec}$ means that the effective scale that is being constrained by the 21-cm power spectrum is close to the pivot scale $k=0.05\,{\rm Mpc}^{-1}$ at which $A_{\rm s}$ is defined~\cite{Planck:2018}.

As we noted in the main text, the form of $C_{\rm PowSpec}$ changes for different scenarios. The coefficients for configuration G are [0.304, 0.0488, 0.484, 0.0698], for configuration B: [0.307, 0.9986, 0.461, 0.0751], for configuration C: [0.311, 1.118, 0.382, 0.0811], for configuration D: [0.315, 1.233, 0.300, 0.0760], and for the CV-only case: [0.335, 2.97, -0.292, 0.0223]. Thus, the coefficients for configurations A and B are nearly identical (since both are strongly dominated by the thermal noise), but things change with C and D (the angular resolution is now higher, and the CV plays some role, particularly for D), and big changes happen for CV-only (as the detailed shape of the power spectrum now plays a major role, and much smaller scales come into play). 

\begin{table}[ht]
\begin{center}
\begin{minipage}{340pt}
\caption{The correlation coefficients in the fits of the 21-cm power spectrum. Note that the actual parameters used in the fitting are the logarithms of the parameters listed here (except for $n_{\rm s}$).}
\label{tab:corr}
\begin{tabular*}{\textwidth}{@{\extracolsep{\fill}}l|cccc|cccc@{\extracolsep{\fill}}}

\toprule

\multicolumn{1}{@{}c@{}}{} & \multicolumn{4}{@{}c@{}}{CV only} & \multicolumn{4}{@{}c@{}}{Configuration A} \\\cmidrule{2-5}\cmidrule{6-9}%

{$\Omega_{\rm b}h^2$} & 0.538 & & & & -0.661 \\

{$\Omega_{\rm m}h^2$} & -0.965 & -0.423 & & & -0.779 & 0.317 \\

{$A_{\rm s}e^{-2\tau}$} & 0.917 & 0.271 & -0.897 & & 0.633 & -0.992 & -0.225 \\

{$n_{\rm s}$}  & 0.814 & 0.260 & -0.911 & 0.674 & {-0.0812} & 0.575 & -0.487 & -0.658 \\

\cmidrule{2-5}\cmidrule{6-9}%

& {$H_0$} & {$\Omega_{\rm b}h^2$} & {$\Omega_{\rm m}h^2$} & {$A_{\rm s}e^{-2\tau}$}
& {$H_0$} & {$\Omega_{\rm b}h^2$} & {$\Omega_{\rm m}h^2$} & {$A_{\rm s}e^{-2\tau}$}\\

\botrule

\end{tabular*}
\end{minipage}
\end{center}
\end{table}

\begin{figure}
\centering
\includegraphics[width=.9\textwidth, angle=0]{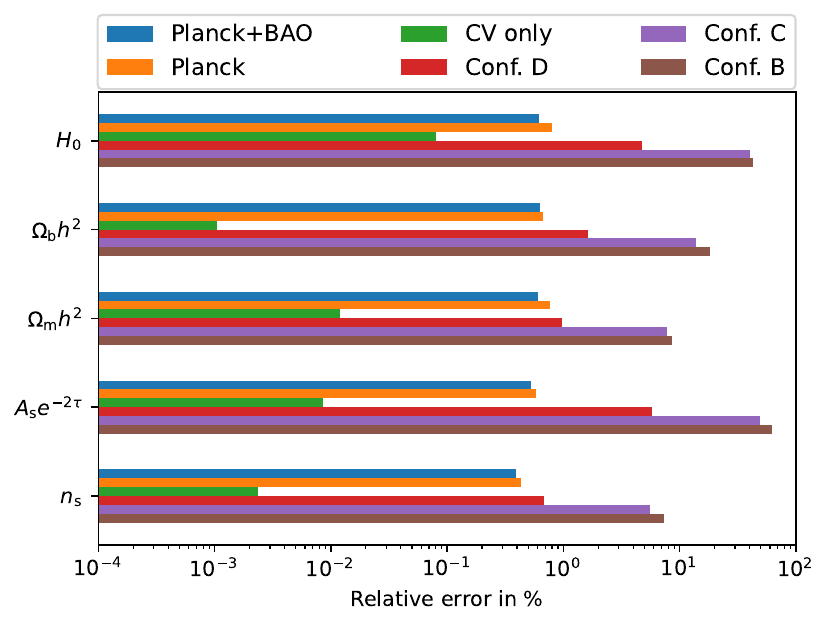}
\caption{The relative 1$\sigma$ errors in \% from fitting to the 21-cm power spectrum from the dark ages. We show graphically the main results listed in Table~\ref{tab:error_power}. Note that `Conf.' stands for configuration.}
\label{fig:relative_error_old}
\end{figure}

In fitting the 21-cm power spectra from the dark ages, in the main text we focused on $C_{\rm PowSpec}$ as well as constraints on Helium and neutrinos. The relative errors in the standard cosmological parameters are listed in Table~\ref{tab:error_power} and shown in Fig.~\ref{fig:relative_error_old}. We do not show configuration A (for which the errors are even significantly larger than for the 1,000\,hr global signal case). For configuration D some of the errors approach Planck levels, while the ultimate CV-only case is in principle better than Planck by between 1 and 3 orders of magnitude. 


\begin{table}
\begin{center}
\caption{For the 21-cm power spectrum, the relative 1$\sigma$ errors in \%, compared to Planck. Note that we include the CV-only case (which has an extra factor of $10^{-2}$ as indicated). We also list here the errors on $\Omega_{\rm c}h^2$ since this is one of the standard input parameters in \texttt{CAMB}. }
\label{tab:error_power}

\begin{minipage}{8.4cm}
\begin{tabular*}
{\textwidth}{@{\extracolsep{\fill}}lcccccc@{\extracolsep{\fill}}}
\toprule

{} & {Planck} & {Planck} & {CV only} & \multicolumn{3}{@{}c@{}}{\underline{~~~~~Configurations~~~~~}}\\

{} & {$+$\,BAO} & & {($\times 10^{-2}$)} & D & C & B \\

\midrule

{$H_0$} & $0.621$ & 0.802 & 8.10 & 4.76 & 40.4 & 42.4 \\

{$\Omega_{\rm b}h^2$} & 0.624 & 0.671 & 0.105 & 1.62 & 13.8 & 18.4 \\

{$\Omega_{\rm m}h^2$} & 0.611 & 0.769 & 1.21 & 0.968 & 7.85 & 8.60 \\

{$A_{\rm s}e^{-2\tau}$} & 0.532 & 0.584 & 0.859 & 5.82 & 49.3 & 62.9 \\

$n_{\rm s}$ & $0.393$ & 0.435 & 0.237 & 0.687 & 5.58 & 7.32 \\

\midrule

{$\Omega_{\rm c}h^2$} & 0.762 & 1.00 & 1.46 & 1.09 & 8.73 & 9.69 \\

\botrule
\end{tabular*}

\end{minipage}
\end{center}
\end{table}


Finally, we explore the dependence of our power spectrum results on varying the assumed observational ranges, for configuration A. For $k$ this is of interest since observational limitations (such as foreground removal) could limit the available range. Table~\ref{tab:error_k} shows that the results are insensitive as long as we include the scales around the first few BAO, where the S/N is maximized.
We also explore various $\nu$ ranges, keeping $\Delta \nu=5$~MHz and removing low redshifts. This is interesting since in rare models the 21-cm signal can be affected by galaxies at redshifts almost up to 35~\cite{subtle}, plus it is of interest to understand to what degree the lower redshifts dominate the fitting. As shown in Table~\ref{tab:error_nu}, since the S/N is maximized at the lowest redshift, the cutoff redshift indeed has a substantial effect on the results; the minimum redshifts corresponding to the tabulated cases are 30, 33.8, 38.7, and 45.1. The precise high-redshift cutoff is less important given the low S/N at that end.


\begin{table}[ht]
\begin{center}
\begin{minipage}{\textwidth}
\caption{The relative (1$\sigma$) errors in \%, for various $k$ ranges, when fitting the power spectrum with configuration A. In all cases we maintain an integer number of bins with $\Delta \ln k=0.5$.}
\label{tab:error_k}
\begin{tabular*}
{\textwidth}{@{\extracolsep{\fill}}lcccc@{\extracolsep{\fill}}}
\toprule%

& \multicolumn{4}{@{}c@{}}{$k$ range [${\rm Mpc}^{-1}$]}\\
\cmidrule{2-5}
{} & {Fiducial [0.00779 - 1.91]} & {[0.0234 - 2.10]} & {[0.0779 - 2.58]} & {[0.00779 - 5.18]}  \\

\midrule

$C_{\rm PowSpec}$ & 4.59 & 4.65 & 7.76 & 4.59  \\

\botrule
\end{tabular*}
\end{minipage}
\end{center}
\end{table}


\begin{table}[ht]
\begin{center}
\caption{The relative (1$\sigma$) errors in \%, for various $\nu$ ranges, when fitting the power spectrum with configuration A. In all cases we maintain an integer number of bins with $\Delta \nu=5$~MHz.}
\label{tab:error_nu}

\begin{minipage}{11.1cm}
\begin{tabular*}{\textwidth}{@{\extracolsep{\fill}}lcccc@{\extracolsep{\fill}}}
\toprule%

& \multicolumn{4}{@{}c@{}}{$\nu$ range [MHz]}\\
\cmidrule{2-5}
{} & {Fiducial $[5.81 - 45.81]$} & {$[5.81 - 40.81]$} & {$[5.81 - 35.81]$} & {$[5.81 - 30.81]$}\\

\midrule

$C_{\rm PowSpec}$ & 4.59 & 6.57 & 12.2 & 32.7 \\

\botrule
\end{tabular*}
\end{minipage}
\end{center}
\end{table}

\subsection{Discussion of foregrounds}
\label{sec:foregrounds}
The brightness temperature of the foreground sky emission at $z=40$ is expected to be around 13,070\,K. While this is significantly higher than at lower redshifts, the thermal noise is proportional to the sky brightness for the global signal (eq.~\ref{eq:thermal_global}), and the square of the sky brightness for the power spectrum (eq.~\ref{eq:thermal_PS}); thus, the relative accuracy needed for foreground removal, in order for the foreground residuals to fall below the thermal noise, is independent of redshift (for a fixed integration time and frequency bin size). For example, for the global signal with $t_{\rm int}$ = 1,000 hrs, the foreground must be removed to an accuracy of a part in $10^6$ or better (depending on the frequency bin size). This is challenging, but the cosmic dawn experiments are making steady progress, and as explained in the introduction of the main text, we expect the lunar environment to make this task significantly easier than for the terrestrial environment.  

We wish to account for foreground removal while fitting the global signal, at least in the best-case scenario. Thus we add a free parameter $A$ to the  model that we fit to the data, in the shape of the synchrotron foreground, i.e., $A\, \nu^{-2.6}$. In practice, in current global signal experiments more polynomial terms are usually added for a more realistic foreground modeling. As we noted, there are reasons to hope that less of this will be required on the moon, but even in the best-case scenario, a signal component of the same shape as the foreground cannot be distinguished from it. To illustrate the impact, we note that the error in $C_{\rm Global}$ for $t_{\rm int}=1$,000\,hrs, which is 10.1\%, would instead be 5.54\% without this additional foreground term.

In the case of the 21-cm power spectrum, in addition to foreground removal, which can never be perfect, another method of dealing with foregrounds is to avoid them. Foreground contamination is expected to be largely restricted to within a wedge-shaped region in the 2D $(k_\perp, k_{\parallel})$ Fourier space, where these are the components of the wavevector perpendicular and parallel to the line of sight, respectively. Thus, it may be easier to achieve an extremely high accuracy of foreground removal outside the wedge. For a rough estimate of the effect of foreground avoidance, we consider this foreground wedge with different levels of contamination. We calculate the wedge boundary using \cite{Datta2010,dillon14,pober14,jensen15}
\begin{equation}
k_{\parallel} = \left( \frac{r_{\nu} \sin{\theta_{\rm L}}}{r_{\nu}^{\prime} \nu} \right) k_{\perp} \, ,
\end{equation}
where $r_{\nu}$ is the comoving distance to the bin center, $r_{\nu}^{\prime} = \frac{dr}{d\nu}$ at the bin center, and $\theta_{\rm L}$ is the angle on the sky with respect to the zenith from which the foregrounds contaminate the power of the 21-cm signal. At $z=40$, we assume two scenarios. The first assumes $\theta_{\rm L} = 2\times$\,FWHM (optimistic). We estimate that roughly $1/10$ of the $(k_\perp, k_{\parallel})$ space is affected in this case. Assuming a more pessimistic case of $\theta_{\rm L} = \pi/2$, we find that roughly 1/2 of the S/N can be lost due to foreground contamination. Thus, the effect of foreground avoidance can be significant but is most likely not a game changer.


\end{appendices}


\bibliography{refs}


\end{document}